\documentclass[a4paper]{article}

\usepackage{18lomcon}        
\usepackage{cite}             
\usepackage{epsfig}           
\usepackage{epstopdf}

\usepackage{graphicx}
\usepackage{amsmath,amssymb}

\newcommand{\ee}{\end{equation}}
\newcommand{\be}{\begin{equation}}
\newcommand{\bea}{\begin{eqnarray}}
\newcommand{\eea}{\end{eqnarray}}

\begin{document}


\title{CURRENT STATUS OF WARM INFLATION}

\author{Raghavan Rangarajan \email{raghavan@prl.res.in}
        }

\affiliation{Theoretical Physics Division,
Physical Research Laboratory, Navrangpura, Ahmedabad 380009,
India}


\date{}
\maketitle


\begin{abstract}

Warm inflation is an inflationary scenario in which a thermal bath coexists with the inflaton
during inflation.  This is unlike standard cold inflation in which the Universe is effectively 
devoid of particles during inflation.  The thermal bath in warm inflation
is maintained by the dissipation of the inflaton's
energy through its couplings to other fields.  Many models of warm inflation have been proposed and 
their predictions have been compared with
cosmological data.  Certain models of inflation that are disallowed 
in the context of cold inflation
by the data 
are allowed in the warm inflationary scenario, and vice versa.
\end{abstract}

\section{Introduction}

In this brief article we shall provide a review of warm inflation and its
current status.  We shall first discuss what is warm inflation and how it
is different from the standard cold inflation.  We shall then discuss how
to construct a warm inflation model.  Finally we shall consider the compatibility of various
warm inflation models with the cosmic microwave background data.

\begin{figure}[h!]
\includegraphics[width=7cm]{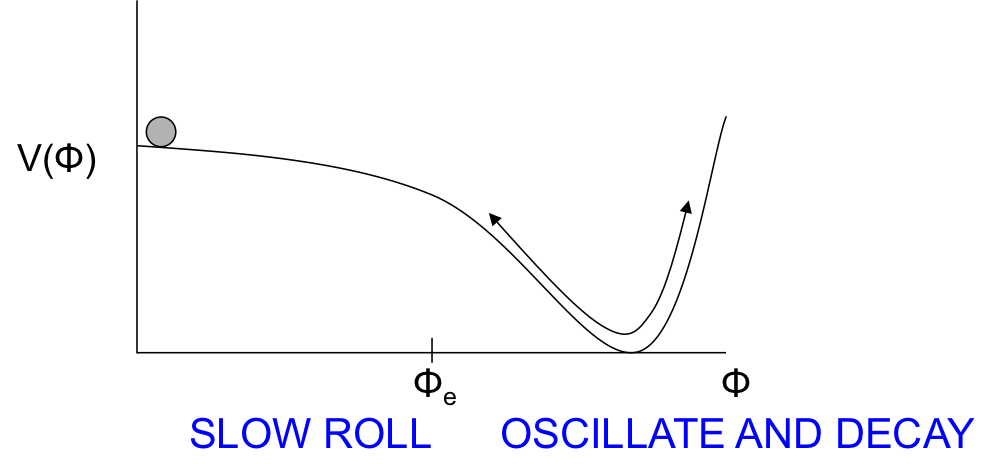}
%
%
%
\caption{A cartoon of the inflaton potential.}
\label{fig:inflaton_potential_phi}
\end{figure}

Inflation is a period of accelerated expansion in the early Universe that
occurred when the Universe was $10^{-38}$ s old or later.  It is invoked to solve
the horizon and flatness problems.  As a bonus, it provides a mechanism 
for generating the primordial energy density perturbations that are the 
seed for late time structure formation (which starts at 
$t\sim$ 70,000 years.  During inflation the energy density of the Universe is dominated by the potential energy of a slowly moving
scalar field $\phi$ called the inflaton.  In Fig. \ref{fig:inflaton_potential_phi} we see
a cartoon of the inflaton potential.  For $\phi<\phi_e$ the potential is flat
and the field rolls slowly.  For $\phi>\phi_e$
the field oscillates in its potential and decays.

In an expanding Universe the scale factor $a(t)$ indicates how the physical distance $d$
between points in space scales with time, $d(t) \propto a(t)$.  During inflation $a$ increases as $\exp(Ht)$,
where $H$ is the Hubble parameter during inflation.  $Ht$ increases by
a factor of at least 60 (for GUT scale inflation) and so any finite volume
in the Universe increases by a factor of $\exp(180)$.  (An increase in the scale factor by $e^N$ is referred to as there being $N$ e-foldings of inflation.)  Therefore
the number density of any species goes to practically 0 leaving the 
Universe in a supercooled state. After the inflationary era is over the
inflaton decays, its decay products thermalise, and one finally has a 
thermal bath of quarks, leptons, gauge bosons, higgses, dark matter
particles and other Beyond the Standard Model particles.   This
latter phase is called the reheating era.  A key issue in cold inflation
described above is that one ignores any decay of the inflaton during
inflation.

In the warm inflation scenario the Universe inflates as in cold inflation.
However one considers the decay of the inflaton during inflation.  Hence
the number density of particles does not go to 0 during inflation.  If the
dissipation is fast enough so as to maintain a thermal bath with $T>H$
then one has a warm inflation scenario
\cite{BereraFang,Berera:WI}.  In some warm inflation models
there is no need of a separate reheating era.

There are several models of inflation - over 70 single field inflation models
are listed in Encylopedia Inflationaris \cite{martin_encyl}.  So why should
one consider a new scenario like warm inflation? 
Firstly, it is natural to consider the effects of the inflaton couplings not just during
reheating but also during inflation.  (Whether or not one will
get a sufficiently hot thermal bath is a different matter, as we
shall see.). Furthermore, for some warm inflation models, the
eta problem, namely, the presence of large quantum
corrections to the inflaton potential that ruins its flatness, is 
resolved.  Also, some potentials that are excluded by
cosmic microwave background (CMB) data in the cold inflation scenario are allowed in the
warm inflation scenario (though the converse is also true).

\section{How is Warm Inflation Different from Cold Inflation?}

It may be noted that
warm inflation constitutes a different paradigm of inflation.  The presence of a thermal bath differentiates it from the cold inflation scenario.  While studying inflation one considers
the homogeneous background field $\phi(t)$ and its spatial
perturbations $\delta\phi({\bf{x}},t)$, or their Fourier transform,
$\delta\phi_k(t)$.  Both the background field and the perturbations
are affected by the presence of the thermal bath.

We first consider the homogeneous inflaton field $\phi(t)$.  The equation
of motion of this background field is given by
\be
\ddot\phi + (3H +\Gamma)\dot\phi +\frac{dV}{d\phi}=0\,,
\ee
where $\Gamma$ is a dissipation coefficient due to inflaton couplings
to other fields,  which is not considered in
cold inflation during the inflaton slow roll phase.  When $\Gamma>H$ it helps
to slow down the inflaton.  
The slow roll parameters for warm inflation are
given by
\begin{equation}\label{slow_roll_parameters}
\epsilon = \frac{M_{Pl}^2}{16\pi}\,\left(\frac{V_\phi}{V}\right)^2, \quad \eta = 
\frac{M_{Pl}^2}{8\pi}
\,\frac{V_{\phi\phi}}{V},\quad\beta = 
\frac{M_{Pl}^2}{8\pi}
\,\left(\frac{\Gamma_{\phi}\,V_\phi}{\Gamma\,V}\right)\,
\end{equation}
where $M_{Pl}=1.2\times 10^{19}$ GeV is the Planck mass.
The slow roll conditions needed for the inflationary phase are
\begin{equation} \label{slow_roll}
\epsilon \ll 1+Q,\quad |\eta| \ll 1+Q,\quad |\beta| \ll 1+Q\,,
\end{equation}
where $Q=\Gamma/(3H)$.  The presence of Q on the right hand side, which
is obviously absent in cold inflation, implies
that the slow roll conditions can be satisfied even if the slow roll parameters are 
large, if $Q\gg 1$, as it happens in some models of warm inflation.  In these
models of warm inflation, the eta problem is solved.  Finally in some models of
warm inflation, before the slow roll conditions break down the inflaton energy
density becomes smaller than the radiation energy density.  In that case
inflation ends but then there is no separate reheating phase because one has an automatic transition to the radiation
dominated era (though the inflaton
will eventually oscillate and decay).

The thermal bath affects the inflaton perturbations and thereby the primordial curvature
perturbations.   The 
curvature
perturbations affect the CMB anisotropy and the large scale structure that we observe today. The equation of motion for the inflation perturbations
in the presence of the thermal bath is given by \cite{Hall:2003zp,Ramos:2013nsa,Bartrum:2013fia}
\begin{equation} \label{eq_motion_perturbation}
\delta\ddot{\phi}_k + (3H+\Gamma)\,\delta\dot{\phi}_k + \left(\frac{k^2}{a^2} + V_{\phi\phi}\right)\delta \phi_k = \sqrt{2\,\Gamma\,T}\,a^{-3/2}\,\xi_k\,,
\end{equation}
where $\xi_k$ represents thermal noise.  The above is a form of the Langevin equation with the fluctuation term on the r.h.s. related to the dissipation term on
the l.h.s.  The primordial curvature power spectrum is proportional to
$|\delta\phi_k|^2$ (in the spatially flat gauge), where $\delta\phi_k$ is evaluated when the physical wavelength of the perturbation $(\lambda_{phys}=2\pi a(t)/k )$ becomes large enough
that $\delta\phi_k$ becomes constant,
or freezes out \cite{Berera:1999ws}.  

We are concerned only with the perturbations that 
correspond to cosmologically relevant length scales today, from $10^{-3}$ Mpc to
14000 Mpc \cite{Lyth:2009zz}.
This corresponds to about 16 e-foldings of 
inflation, starting from about 60 e-foldings of inflation before the end of 
inflation (for GUT scale inflation).  The observed CMB anisotropy reflects perturbations on scales of 10 Mpc and larger.
 When
 $Q\ll 1 \,(Q\gg1)$ 
it is referred to as weak (strong) dissipative
warm inflation.  
The inflaton perturbations for cold inflation, weak dissipative warm inflation and strong dissipative warm inflation are given by
\cite{Berera:2006xq,Ramos:2013nsa}
\begin{align}
\delta\phi_k &\sim  H &&\textrm {Cold Inflation}\nonumber\\
\delta\phi_k &\sim  \sqrt{H T} && \textrm {Weak Dissipative Warm Inflation}\nonumber\\
\delta\phi_k &\sim  \sqrt{T(H\Gamma)^\frac12} && \textrm {Strong Dissipative Warm Inflation}
\label{coldwarmpert}
\end{align}
The primordial curvature power spectrum (or scalar power spectrum)
is given in 
Ref. \cite{Bartrum:2013fia} 
(based on Refs. \cite{BereraFang, Moss:1985wn, Berera:1999ws, Hall:2003zp, Ramos:2013nsa})
as 
%
\begin{equation} \label{power_spectrum}
P_\mathcal{R}(k)=\left({H_k\over\dot\phi_k}\right)^2\left({H_k\over 2\pi}\right)^2\left[1+2n_k+\left(T_k\over H_k\right){2\sqrt{3}\pi Q_k\over\sqrt{3+4\pi Q_k}}\right]~,
\end{equation}
where the subscript $k$ indicates that the variable is evaluated at the time of
horizon crossing of the $k$ mode perturbation $\delta\phi_k$, and $n_k$ 
represents the distribution of inflaton particles in the thermal bath.  In the literature, one considers either $n_k=0$ or the Bose-Einstein
distribution, $n(k)=[\exp\{k/(aT)\}-1]^{-1}$.  
For the latter case, $1+2n_k=\coth[H_k/(2T_k)]$, using $k/a_k=H_k$.
In addition to the explicit temperature dependence in the square bracket 
above, the prefactor (whose form is the same as that for cold inflation) will
reflect the influence of dissipation.  
Note that $[H_k/(2\pi)]^2$ times
the first term in the square bracket reflects the standard quantum
contribution, as in cold inflation, its product with 
$\coth[H_k/(2T_k)]$ reflects the weak dissipative warm inflation result
for $T\gg H$, and
the product with the last term indicates
the 
strong dissipative warm inflation result, as in
Eq. (\ref{coldwarmpert}).

Inflation gives rise to both scalar and tensor perturbations of the metric.
Gravitational waves are weakly coupled to the thermal bath and so the tensor
power spectrum has the same form as in cold inflation, namely,
\be
P_T(k)=\frac{16}{\pi} \frac{H_k^2}{M_{Pl}^2}\,.
\ee
The tensor-to-scalar ratio $r$ is defined, as usual, as
\be
r=\frac{P_T(k_P)}{P_\mathcal{R}(k_P)}
\ee
where $k_P$ refers to the pivot scale, a fiducial scale for which 
there is greater observational accuracy for any particular experiment.

\section{Constructing a Warm Inflation Model}

The dissipation coefficient $\Gamma$ reflects the transfer of energy from
the inflaton field to the thermal bath.  If one couples the radiation, i.e. light fields
with mass $m<T$, directly to the inflaton then one gets a $\Gamma\dot\phi$
term in the equation of motion for $\phi$ but one also gets large thermal
corrections to the inflaton potential and too few e-foldings of inflation
\cite{BereraGleiserRamos1998,yokoyamalinde1999}.  There are two approaches
to avoiding this.  In the first approach one couples the inflaton only to heavy
fields through a superpotential of the form \cite{Moss:2006gt}
\begin{equation} \label{superpotential}
W=f(\Phi)+{g\over2}\Phi X^2+ {h\over 2}XY^2~,
\end{equation}
where $\phi$ is associated with the scalar component of the $\Phi$ superfield,
and the $X$ fields are heavy, i.e $m_X>T$, while the $Y$ fields are light 
$(m_Y<T)$.
The inflaton field can decay to $Y$ particles either through virtual $X$ when
$T\ll m_X$, or
through  decay to real $X$ which then decay to $Y$ when $T<m_X$ and
$h\sqrt{N_Y}$ is small \cite{BBRR2013}.  The form of the dissipation
coefficient is $\Gamma=C_\phi T^3/\phi^2$.  
The heavy $X$ ensure that the 
thermal corrections are small and supersymmetry ensures that the vacuum
corrections are small.  However, viable models of warm inflation need
$10^6$ or $10^4$ $X$ fields to satisfy warm inflation requirements, particularly
$T>H$ during inflation \cite{Berera:1998px,Bastero-Gil:2015nja}.
Such a large number of fields are obtained by considering brane-antibrane 
models of inflation where the $X$ fields correspond to strings stretched between
brane and antibrane stacks
\cite{BasteroGil:2011mr}, or extra-dimensional scenarios with a tower of 
Kaluza-Klein modes \cite{Matsuda:2012kc}.

In the second approach to constructing a warm inflation model, one makes the inflaton field a pseudo-Nambu-Goldstone boson.  This has been realised in warm natural inflation models
\cite{Mishra:2011vh},
and the warm
little inflation model (which is similar to the little Higgs model) 
\cite{Bastero-Gil:2016qru}.  
In these models it is sufficient for the inflaton 
to couple to a few fields.  In Ref. \cite{Mishra:2011vh}, there is one
additional pseudo-Nambu Goldstone boson besides the inflaton
and another light field,
and $\Gamma\sim\dot\phi^2 T$ and $Q\gg1$. 
In Ref. \cite{Bastero-Gil:2016qru} the inflaton field is coupled to two
light fields, and $\Gamma=C_T T$ and both weak and strong dissipative warm inflation
scenarios are considered.

\section{Comparing with Data}

Various models of warm inflation, as identified by the inflaton potential and
the form of the dissipation coefficient, have been studied and compared
with the cosmological data.  $\Upsilon=C_\phi T^3/\phi^2$ and $\Gamma=C_T T$ are the 
usual forms of the dissipation coefficient considered in the literature.  In general, 
$\Gamma = C_\phi T^c \phi^{2a}/M_X^{2b}$
with $c+2a-2b=1$ \cite{Ramos:2013nsa}. 

\begin{figure}[h!]
\includegraphics[width=7cm]{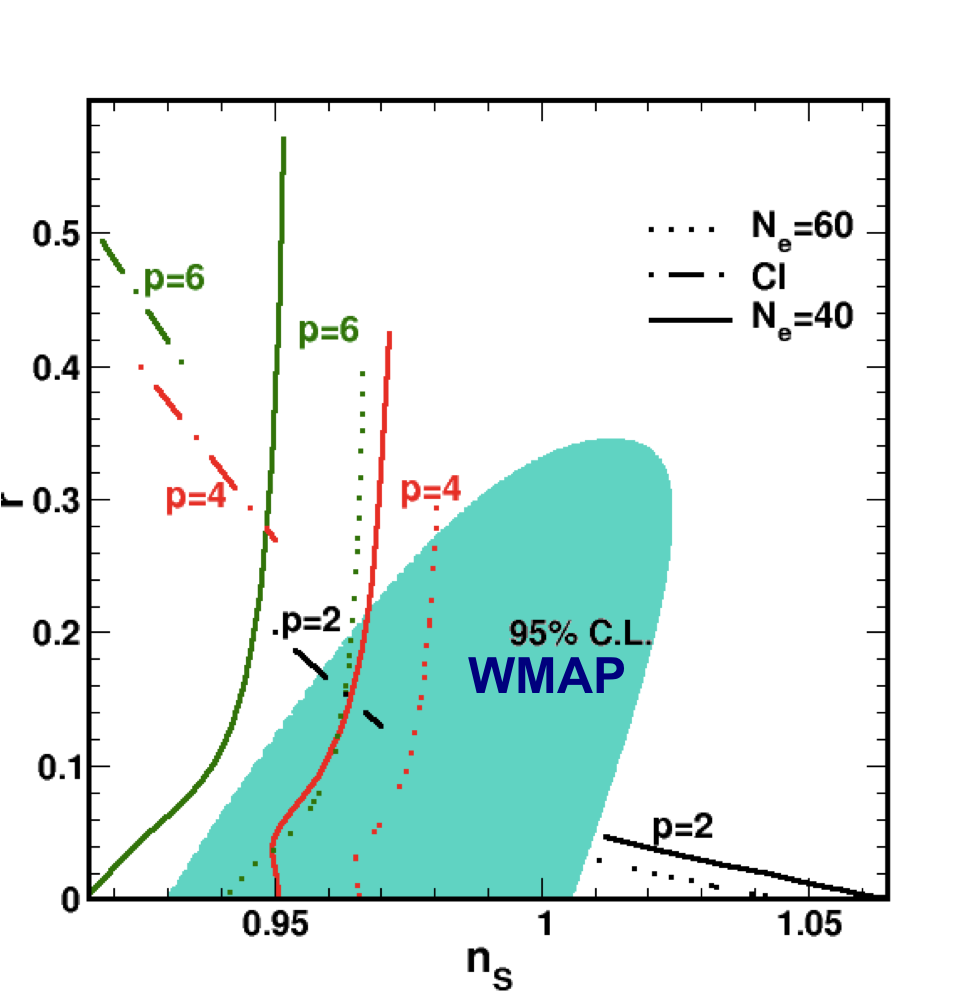}
%
%
%
\caption{Allowed region by WMAP in $r-n_s$
plane is shown in teal.  Also plotted are the $r-n_s$ values obtained for quadratic, quartic
and sextic warm inflation models (solid and dotted lines) 
for $N_e$=40 and 60, and cold inflation models (dot-dash) for varying $N_e$.}
\label{fig:Mar-WMAP-IJMP2009-Helsinki}
\end{figure}

Limits from cosmological data are often written in terms of allowed values of
the spectral index $n_s$ defined as 
$n_s-1=
d \ln P_\mathcal{R}/d \ln k|_{k_P}$
and the tensor-to-scalar ratio $r$.
In Fig. \ref{fig:Mar-WMAP-IJMP2009-Helsinki}
one sees the region in the $r-n_s$
plane allowed by WMAP in teal \cite{BasteroGil:2009ec,MarTalkHelsinki2013}.  Also plotted are the $r-n_s$ values obtained for warm inflation models with monomial potentials $(V\sim\phi^n)$  as
separate curves in the figure.  $n=2,4$ and 6 are considered, for
two values of the number of e-foldings of inflation
from when the perturbation associated with the pivot scale crosses the horizon till the end of inflation, i.e. $N_e$ equal to
60 and 40.  We can focus on the $N_e=60$ 
curves for warm inflation.   Along each curve, the different points 
correspond to different values of $Q(k_P)$ with the values increasing as 
one goes down the curve.  The cold inflation curves (CI) are also
shown.  For cold inflation models the different points
correspond to values of $N_e$ varying from 50 to 60 (from left to right).  

We notice that quadratic warm inflation has too large a value of $n_s$
and so is disallowed, while it is consistent with the data for cold inflation.
On the other hand, quartic and sextic
cold inflation are ruled out by the data while they 
are allowed in warm inflation for appropriate values of $Q(k_P)$.

\begin{figure}[h!]
\includegraphics[width=7cm]{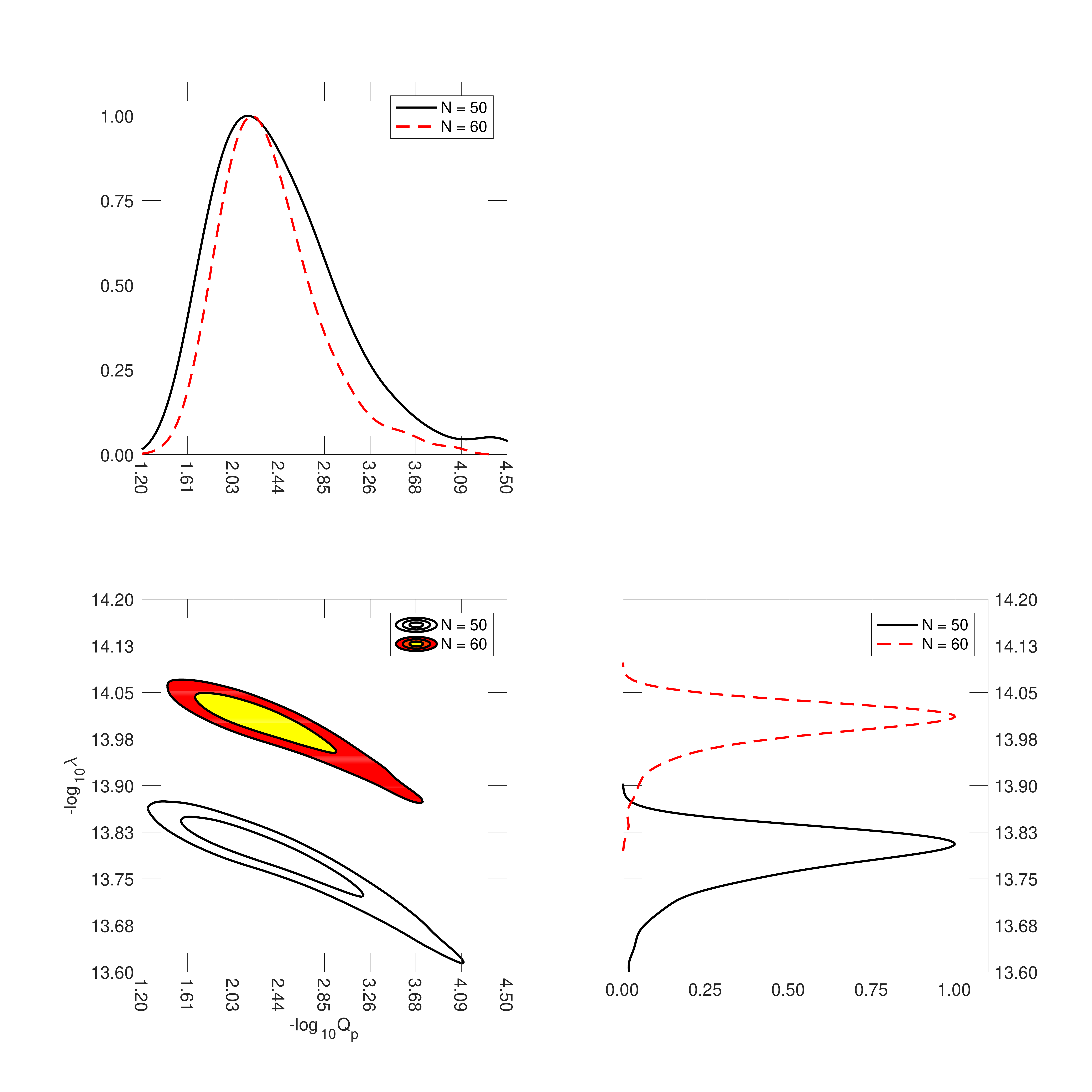}
%
%
%
\caption{The joint probability distribution for the inflaton self-coupling
$\lambda$ and $Q(k_P)$ for $N_e=50$ and 60.  
%
The preferred range of values for $\lambda$ is
$1.5\times10^{-14} $  to  $1.9\times10^{-14}$  
for $N_e=50$, and is $9.2\times10^{-15}$   to   $1.1\times10^{-14}$ for $N_e=60$.
The preferred range of values for $Q_P$ is 
$9.5\times10^{-4}$   to $1.4\times10^{-2}$  for
$N_e=50$, and is $1.6\times10^{-3}$  to $1.2\times10^{-2}$ for $N_e=60$.
}
\label{fig:adgpr2017-jointfig}
\end{figure}

In Ref. \cite{Benetti:2016jhf} the authors perform a Markov Chain
Monte Carlo analysis of the parameters of warm inflation using
the publicly available CosmoMC programme \cite{cosmomc}
and the Planck data.  
They perform this analysis for quartic, sextic, hilltop, Higgs and plateau
sextic warm inflation models for $\Gamma\propto T^3$ and $T$ 
and find parameters compatible with the
Planck data  for $Q(k_P)$ between
$10^{-3}$ and 1.4 and $r$ between $10^{-9}$ and 0.036 for different models.
Another CosmoMC analysis of quartic warm inflation
using Planck data obtains
the joint probability distribution for the inflaton self-coupling
$\lambda$ and $Q(k_P)$ for $N_e=50$ and 60 \cite{adgpr2017},
as shown in 
Fig. \ref{fig:adgpr2017-jointfig}.  
From the marginalised distributions of the parameters of the model 
the preferred range of values for $\lambda$ for $N_e=50$ is
$1.5\times10^{-14} $  to  $1.9\times10^{-14}$  with a mean value of $1.6\times 10^{-14}$, and
the preferred range of values for $N_e=60$ is
$9.2\times10^{-15}$   to   $1.1\times10^{-14}$ with a mean value of $1.0\times10^{-14}$.
The preferred range of values for $Q_P$ is 
$9.5\times10^{-4}$   to $1.4\times10^{-2}$  with a mean value of $3.7\times10^{-3}$ for
$N_e=50$, and
the preferred range of values for $N_e=60$ is
$1.6\times10^{-3}$  to $1.2\times10^{-2}$ with a mean value of $4.4\times10^{-3}$.  
Another CosmoMC analysis for a quartic inflaton potential with $\Gamma\propto T$  
has been carried out in Ref. \cite{Bastero-Gil:2017wwl}.

Warm natural inflation models too have been compared with the cosmological data.  In 
the model studied in Ref. \cite{Visinelli:2011jy} it is found that
warm natural inflation is viable for the scale of symmetry breaking 
(that creates the pseudo-Nambu-Goldstone  boson) to be between
the GUT scale and the Planck scale, while 
Ref. \cite{Mishra:2011vh} finds that the symmetry breaking scale 
in their model should be the GUT scale.  In both models what is 
significant is that symmetry breaking scales well below the Planck
scale are allowed.  Planck scale symmetry breaking was one of 
the less attractive features of cold natural inflation.

Ref. \cite{Bastero-Gil:2013owa} shows that hybrid inflation, which
involves the interplay of two fields during inflation, is consistent 
with the data for warm inflation, in contrast with cold inflation.
The viability of brane inflation, G(alileon) inflation and non-canonical inflation
has also been studied in 
Refs. \cite{Cid:2007fk,delCampo:2007cia,Herrera:2017qux}.

\section{Conclusion}
In summary, warm inflation is a viable paradigm of inflation.
Various warm inflation models are compatible with cosmological
data.  Models such as monomial quartic and sextic warm 
inflation and hybrid warm inflation are allowed by the data
while the corresponding models in the cold inflationary scenario
are disallowed by the data.  On the contrary, the quadratic inflationary 
model is disallowed in warm inflation while consistent with the data 
for cold inflation. In the case of natural warm inflation
the relevant energy scale can be brought down from the Planck
scale, as in cold inflation, to the GUT scale.  While the requirement of a large number
of fields coupled to the inflaton to satisfy the conditions for warm
inflation is unattractive this issue has been resolved in warm
inflation models where the inflaton is a pseudo-Nambu-Goldstone
boson.  

There have been interesting results associated with viscosity in 
the thermal bath during inflation, and on the generation of
non-Gaussian fluctuations during warm 
inflation, which have not been
discussed here.


\end{document}